\documentclass[preprint,showpacs,amsmath,amssymb]{revtex4}
\usepackage{graphicx}
\usepackage[]{caption}
\usepackage{amsmath}
\usepackage{amssymb}

\usepackage{graphicx}
\usepackage{dcolumn}
\usepackage{bm}
\usepackage[usenames]{color} 
\def\be{\begin{equation}}
\def\ee{\end{equation}}
\def\bea{\begin{eqnarray}}
\def\eea{\end{eqnarray}}

\begin{document}

\title{Evolution of Gravitational Perturbations in Non-Commutative Inflation}

\author{Seoktae Koh} \email[email: ]{skoh@hep.physics.mcgill.ca} 
\author{Robert H. Brandenberger} \email[email: ]{rhb@hep.physics.mcgill.ca}

\affiliation{Physics Department, McGill University, Montr\'eal, Q.C., 
H3A 2T8, Canada}         


\begin{abstract}
We consider the non-commutative inflation model of \cite{Joao2}
in which it is the unconventional dispersion relation for regular
radiation which drives the accelerated expansion of space. In
this model, we study the evolution of linear cosmological perturbations 
through the transition between the phase of accelerated expansion
and the regular radiation-dominated phase of Standard Cosmology,
the transition which is analogous to the reheating period in scalar 
field-driven models of inflation. If matter consists of only a
single non-commutative radiation fluid, then the curvature perturbations 
are constant on super-Hubble scales. On the other hand, if we
include additional matter fields which oscillate during the transition
period, e.g. scalar moduli fields, then there can be
parametric amplification of the amplitude of the curvature perturbations.
We demonstrate this explicitly by numerically solving the
full system of perturbation equations in the case where matter
consists of both the non-commutative radiation field and a light scalar
field which undergoes oscillations. Our model is an example where
the parametric resonance of the curvature fluctuations is driven by
the oscillations not of the inflaton field, but of the entropy mode. 
\end{abstract}

\maketitle

\section{Introduction}

Non-commutativity of space and time is a general prediction for an
effective field theory coming from string theory \cite{SSUR}. This
leads to a maximum wavenumber $p_c$ for any field which is subject
to this uncertainty. In turn, this leads
to a dispersion relation which is modified at high densities. In
particular, regular radiation is described by such a modified dispersion
relation \cite{Joao1}. As shown recently 
in \cite{Joao2}, the dispersion relation will likely contain
two branches (two frequencies for any wavenumber smaller than
the maximal momentum). This two-branch structure may be related
to the T-duality symmetry of string theory which has been made
use of in a different context in string cosmology in \cite{BV}.
At high temperatures, both branches of the dispersion relation are
occupied, whereas at low temperatures only the lower branch, the
branch which for $p \ll p_c$ has the regular linear form, is
occupied. 

In \cite{Joao2} it was realized that for suitable classes of
dispersion relations, regular radiation leads to inflationary
expansion of space at high temperatures. The reason is that as
space expands and the physical momentum of a fixed comoving mode
decreases, the energy of the mode increases. This increasing energy
leads (in the context of a background described by Einstein gravity)
to accelerated expansion. We called this model ``non-commutative
inflation".

Recently \cite{Koh} we have studied the generation of cosmological
fluctuations in non-commutative inflation. In contrast to
simple scalar field-driven inflationary models, the fluctuations
are thermal of origin and not quantum vacuum fluctuations, the
reason being that matter in our scenario is a thermal bath rather
than an ultracold Bose condensate of a scalar field. In the limit
that the accelerated expansion is almost exponential, the
spectrum of fluctuations is almost scale-invariant, as should
be expected from the general symmetry arguments of \cite{Press}.

An interesting aspect of non-commutative inflation is that
the transition between the period of accelerated expansion and
the radiation phase of standard cosmology is straightforward:
matter always is in regular radiation. As the universe expands,
the equation of state of this radiation changes from being
that of an accelerated universe early on to that of ordinary
radiation. Thus, there is no need for any period of ``reheating",
a rather non-trivial phase in usual scalar field-driven inflationary
models.
 
In this paper we study the evolution of the fluctuations through
the transition between the inflationary phase and the phase when
the equation of state changes to that of ordinary radiation. Since
there are no oscillations in the equation of state of that matter
giving rise to inflation (in scalar field-driven inflation the
phase of accelerated expansion ends with oscillations of the
scalar field giving rise to inflation which translates into
oscillations in the equation of state of matter). As a consequence,
there is no parametric resonance instability of matter
\cite{TB,DK} or metric \cite{BKM,Fabio1,BaVi,Fabio2} fluctuations in this
phase, at least in the absence of other matter fields which
undergo oscillations.

In models of matter beyond the Standard Model, in particular in models
motivated by superstring theory, there are many scalar matter fields
which are expected to undergo oscillations at the end or after
the period of inflation. These fields lead to entropy modes. Here,
we will study the evolution of cosmological fluctuations in the
presence of such oscillating entropy modes (oscillating ``moduli"
fields). We find that these entropy modes can indeed lead to
parametric resonance of the curvature fluctuations on super-Hubble
scales.

The importance of parametric resonance instabilities in the
process of reheating at the end of inflation was first recognized
in \cite{TB} (see also \cite{DK}). The resonance of matter 
fluctuations during reheating in scalar field-driven inflation
was then studied in more detail in \cite{KLS1} (where the
name ``preheating" for this resonance phenomenon was coined) 
and \cite{STB}. A detailed discussion of the efficient
broad-band and the much less efficient narrow-band resonances
was given in \cite{KLS2}. It was first conjectured in \cite{BKM} that
super-Hubble (but sub-horizon) scales metric fluctuations may
also be parametrically amplified. In particular, there are
no causality constraints \cite{Fabio1} which prohibit this
(recall that in inflationary cosmology the horizon is larger
than the Hubble radius by a factor of $\exp(N_e)$, where $N_e$ is the
number of e-foldings of the inflationary phase). However, as
shown in \cite{Fabio1}, in the case of purely adiabatic fluctuations
there is no resonance at linear order in perturbation theory.
However, it was then established in \cite{BaVi,Fabio2} 
that oscillations in the equation of state of matter
can lead to an increase of the curvature fluctuations on
super-Hubble scales in the presence of low mass (mass smaller
than the Hubble rate during inflation) entropy modes.

In previous work, most of the attention has focused on oscillations
of the inflaton driving the parametric resonance of the entropy
modes of cosmological fluctuations, which in turn feeds the
exponential growth of the curvature perturbation via the usual
sourcing of curvature fluctuations via an entropy inhomogeneity.
In the context of non-commutative inflation, it is the background
value of the entropy field which undergoes oscillations. Here, we
show that such oscillations may also be able to excite a parametric
resonance for entropy fluctuations. 

The outline of this article is as follows: In the following section
we give a brief review of non-commutative inflation. Then, in
Section 3, we study the growth of cosmological fluctuations during
the transition between accelerated expansion and the usual radiation
phase in the case of a single component of matter, the radiation.
As expected, we find constant curvature fluctuation on super-Hubble
scales. The key section of our paper is Section 4, in which we
study the equations of motion for cosmological fluctuations for
non-commutative radiation coupled to an oscillating scalar field
(the modulus field acting as the entropy mode). We derive the
relevant perturbation equations, solve them numerically, and
give some analytic insight into the solutions. We conclude
with a discussion of our results.

\section{Noncommutative inflation}

The starting point of the non-commutative inflation model of
\cite{Joao2} is the modified dispersion relation for massless
particles
\be \label{disprel1}
E^2  - p^2 c^2 f(E)^2 \, = \, 0 \,
\ee
which results from the non-commutativity of space and time. Here,
\be \label{disprel2}
f(E) \, = \, 1 + (\lambda E)^{\alpha} \, .
\ee
Note that $p$ and $E$ denote momentum and energy,
respectively, $\alpha \ge 1$ is a positive constant, and
the length scale $\lambda$ determines 
the maximal momentum $p_c$. For $\alpha > 1$ the dispersion 
relation has two branches. On the upper branch, the energy increases as
the momentum decreases while the
universe is expanding. It is this behavior which, for a range
of values of $\alpha$ explored in \cite{Joao2}, leads to inflationary 
expansion of space when the usual Friedmann-Robertson-Walker 
equations for the coupling of the background space-time to matter
are used.

The modification of the dispersion relation leads to a
deformed thermal spectrum \cite{Joao1}. The energy density $\rho$ is given
by 
\bea \label{edens}
\rho \, = \, \frac{1}{\pi^2}\int \frac{E^3}{e^{E/T}-1}\frac{1}{f^3}
\left|1-\frac{f^{\prime}E}{f}\right| \, ,
\eea
and the expression for the pressure ${\cal P}$ is
\be \label{pressure}
{\cal P} \, = \, 
\frac{1}{3}\int \frac{\rho(E) dE}{1-\frac{f^{\prime}E}{f}} \, ,
\ee
In the high energy limit, $\lambda E \gg 1$, $f^{\prime}E/f \simeq \alpha$
and $\rho \propto T$. This approximation leads to a nearly constant
equation of state parameter $w \equiv {\cal P} / \rho$.

Regular commutative inflation has no intrinsic entropy fluctuations
since
\bea
c_a^2 \equiv \frac{\dot {\cal P}}{\dot \rho}
\, = \, c_s^2 \equiv \frac{\delta {\cal P}}{\delta \rho} \,
\eea
where $\delta$ indicates spatial variations. The same result holds
for non-commutative radiation. This can be seen by taking on one hand
the time derivatives of Eqs. (\ref{edens}) and (\ref{pressure}), 
and on the other hand the infinitesimal spatial variations of these
quantities. The vanishing of the intrinsic entropy fluctuations
will be important later on. 

\begin{figure}
\includegraphics[height=8cm]{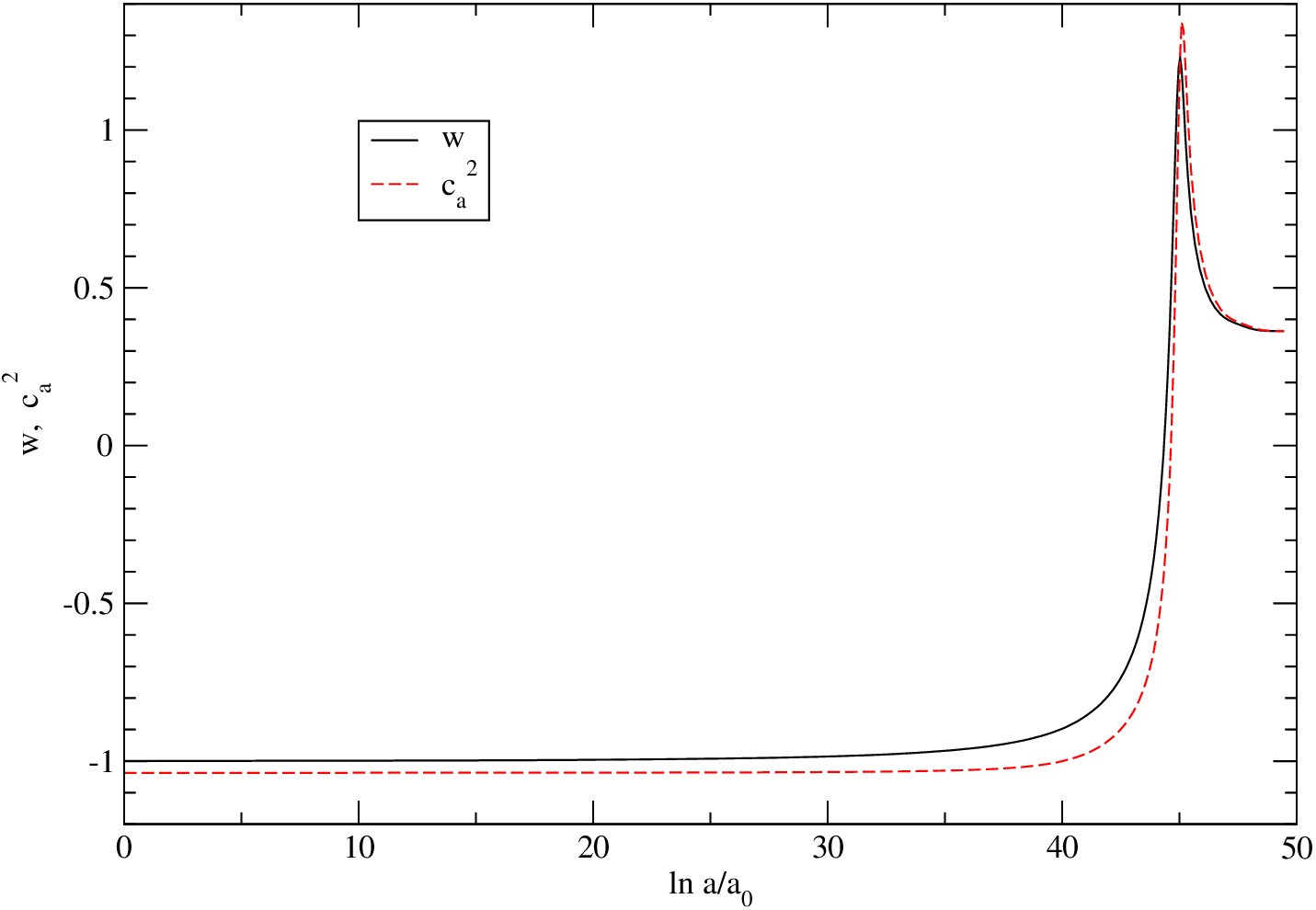}
\caption{Evolution of the equation of state parameter $w$ and of the adiabatic
sound speed $c_a^2$ for a noncommutative fluid during and after
the period of inflation. The value of $\alpha$ is set to $1.099$.
}
\label{fig_eos}
\end{figure}

If the background geometry is described by the Einstein action, the
Friedmann and energy conservation equations take their usual form
\bea
\left(\frac{\dot{a}}{a}\right)^2 &\, \equiv& \, 
H^2 \, = \, \frac{8\pi G}{3} \rho,  \\
\dot{\rho} \, &=& \, -3 H(1+w)\rho \, .
\eea
Inserting the expressions (\ref{edens}) and (\ref{pressure}) into the
above equations, it follows that at high temperatures, a period
of accelerated expansion results provided that the parameter $\alpha$
is suitably chosen \cite{Joao2} (values of $\alpha$ slightly larger
than 1). The resulting evolution of the equation of state parameter
$w$ and of the adiabatic sound speed $c_a^2$ is shown in Fig. \ref{fig_eos}.

Inflation has to last more than 60 number of $e$-foldings (this number
assumes that the energy scale at which inflation takes place is close
to its upper bound) to solve the problems of Standard Cosmology which 
inflation was designed to address.
If noncommutative inflation ends roughly when $\lambda T_{end} \simeq 1$,
then the number $N_e$ of e-foldings is given by
\bea
N_e \, = \, \ln [a(t_{end}) / a(t_i)] \, = \, 
\int^{t_{end}}_{t_i} H dt  \, = \,  
\int_{\lambda T_{end}}^{\lambda T_i} \frac{d\rho/dT}{3(1+w)\rho}dT \, .
\eea
In the high energy limit, we obtain
\bea
N_e \, \simeq \, \frac{1}{3(1+w)}\ln \frac{\lambda T_i}{\lambda T_{end}} \, 
= \, \frac{\ln \lambda T_i}{3(1+w)} \, ,
\label{efold}
\eea
where we have used $\lambda T_{end} \sim 1$.

\section{Evolution of Noncommutative Fluid Perturbations}

\subsection{Equations of motion}

In this section we will take matter to consist only of the non-commutative
fluid. In this case, since there are no entropy fluctuations we expect
the curvature fluctuations to be conserved on super-Hubble scales.
In addition, since there is no oscillating matter field, there cannot be
any parametric resonance effects.

We will work in longitudinal gauge to study the linearized
equations of motion for cosmological perturbations (see \cite{KS,MFB}
for comprehensive reviews of the theory of cosmological perturbations 
and \cite{RHBrev1} for a pedagogical overview). In this gauge, the 
metric takes the form
\bea
ds^2 \, = \, -(1+2\Phi)dt^2 + a^2 (1-2\Psi)\gamma_{ij}dx^i dx^j \, ,
\eea
where $\gamma_{ij}$ is the spatial part of the background metric which
we in the following will take to be Euclidean. The perturbations are
described by $\Phi$ (the Bardeen \cite{Bardeen} potential) and $\Psi$,
both functions of space and time.

If matter consists only of the noncommutative (NC) fluid, the 
perturbed energy-momentum tensor can be written as
\bea
\delta {T^0}_0 = -\delta \rho, \quad
\delta {T^0}_i = \delta q_i, \quad
\delta {T^i}_j = \delta {\cal P} \delta^i_j,
\eea
where we have assumed vanishing anisotropic stress.

The perturbed Einstein equations are
\bea
& & 3H(H\Phi+\dot{\Psi}) - \frac{1}{a^2}\nabla^2\Psi 
\, = \, -4\pi G \delta \rho,
\label{eeq00} \\
& & H\Phi +\dot{\Psi} \, = \, -4\pi G \delta q, \\
& & \ddot{\Psi}+3H\dot{\Psi} + H\dot{\Phi} + (2\dot{H}+3 H^2)\Phi
-\frac{1}{3a^2}\nabla^2(\Psi - \Phi) \, = \, 4\pi G \delta {\cal P}.
\label{eeqij}
\eea
The vanishing of the anisotropic stress tensor makes it possible to
set $\Phi = \Psi$.

The conservation of energy-momentum tensor for the NC fluid leads to
the following equations for $\delta \rho$ and $\delta q$,
\bea
& &\delta \dot{\rho} + 3H(\delta \rho+ \delta {\cal P}) 
- 3\rho(1 + w)\dot{\Phi}
+\frac{1}{a^2} \nabla^2\delta q \, = \, 0, 
\label{ece} \\
& &\delta \dot{q} + 3H \delta q +\rho(1+w)\Phi + \delta p \, = \, 0.
\label{mce}
\eea

Combining Eqs. (\ref{eeq00}) and (\ref{eeqij}) and working in 
momentum space $\nabla^2 \Phi\rightarrow -k^2\Phi_k$, we can obtain a single
differential equation
\bea
\ddot{\Phi}_k + (4+3c_s^2)H\dot{\Phi}_k +\left(\frac{k^2 c_s^2}{a^2}
+2\dot{H}+3H^2(1+c_s^2)\right)\Phi_k \, = \, 0 \, ,
\label{eq_bardeen}
\eea
where we have used $\delta {\cal P} = c_s^2 \delta \rho$.

For adiabatic fluctuations, and in particular in our single perfect fluid 
system, $c_s^2 = c_a^2$.
However, in non-adiabatic fluids, for example systems with several fluids or
scalar fields, there may exist intrinsic isocurvature perturbations
which are denoted by
\bea
p \Gamma \,= \, (c_s^2 - c_a^2)\delta \rho.
\eea

The Sasaki-Mukhanov variable \cite{Mukh,Sasaki} for a fluid 
in terms of which the action for cosmological perturbations 
has canonical form, and which is thus useful for the quantization 
of these fluctuations, can be written as $v = a c_s^{-1} Q_f$ where
\cite{MFB, Hwang05}, 
\bea
Q_f \, = \, \frac{1}{\sqrt{\rho+{\cal P}}}
\left(\delta q -\frac{\rho+{\cal P}}{H}\Phi\right).
\label{def_qf}
\eea
If matter is a scalar field, this gauge-invariant quantity becomes  
\bea
Q_{\phi} \, = \, \delta \phi  - \frac{\dot{\phi}}{H}\Phi.
\label{def_qphi}
\eea

In usual scalar field-driven inflationary models, 
the scalar field begins to oscillate
about the minimum of its potential after the end of inflation. These
oscillations lead to a singularity if Eq. (\ref{eq_bardeen}) is used
because of the $1/\dot{\phi}$ terms in the coefficients.
To avoid this singularity during the preheating phase, the
Sasaki-Mukhanov variables must be used instead of $\Phi$ \cite{Taruya97}.
Even in the absence of oscillations, if the change in the equation of
state of the background in smooth, it is safer to use the Sasaki-Mukhanov
variables instead of $\Phi$ (see e.g. \cite{BST,BK}).
 
The situation, however, is different in our fluid-driven inflationary model.
The speed of sounds $c_s^2$ is negative during the inflationary period.
Then, after the end of inflation, $c_s^2$ crosses zero and
evolves to its final positive value $c_s^2 = 1/3$ (see Fig. \ref{fig_eos}).  
The evolution equation for $Q_f$, Eq. (\ref{qeq_ncs}), contains
$(c_s^2)^{-1}$ terms which diverge when $c_s^2 = 0$. Therefore it is 
convenient to use the $\Phi$ equation of motion (\ref{eq_bardeen}) to
evolve the fluctuations in our fluid-driven inflation model in order 
to avoid the singularity at $c_s^2 =0$.  
 
The curvature perturbation $\zeta$ on uniform energy density hyper-surfaces
is related to the Sasaki-Mukhanov variable as
\bea
\zeta \, = \, \Phi - \frac{H}{\dot{H}}(\dot{\Phi}+H\Phi) \,
= \, -\frac{H}{\sqrt{\rho+{\cal P}}}Q_f.
\label{def_zeta}
\eea
The evolution equation for $\zeta_k$ is
\bea
\dot{\zeta}_k \, = \, \frac{H}{\dot{H}}\frac{k^2 c_s^2}{a^2}\Phi_k
- \frac{H}{\rho(1+w)}p\Gamma.
\label{eq_zeta}
\eea
In the case of adiabatic fluids when ($p\Gamma = 0$), then for long-wavelength
perturbations $\zeta_k$ is conserved. 

\subsection{Numerical results}

\begin{figure}
\includegraphics[height=8cm]{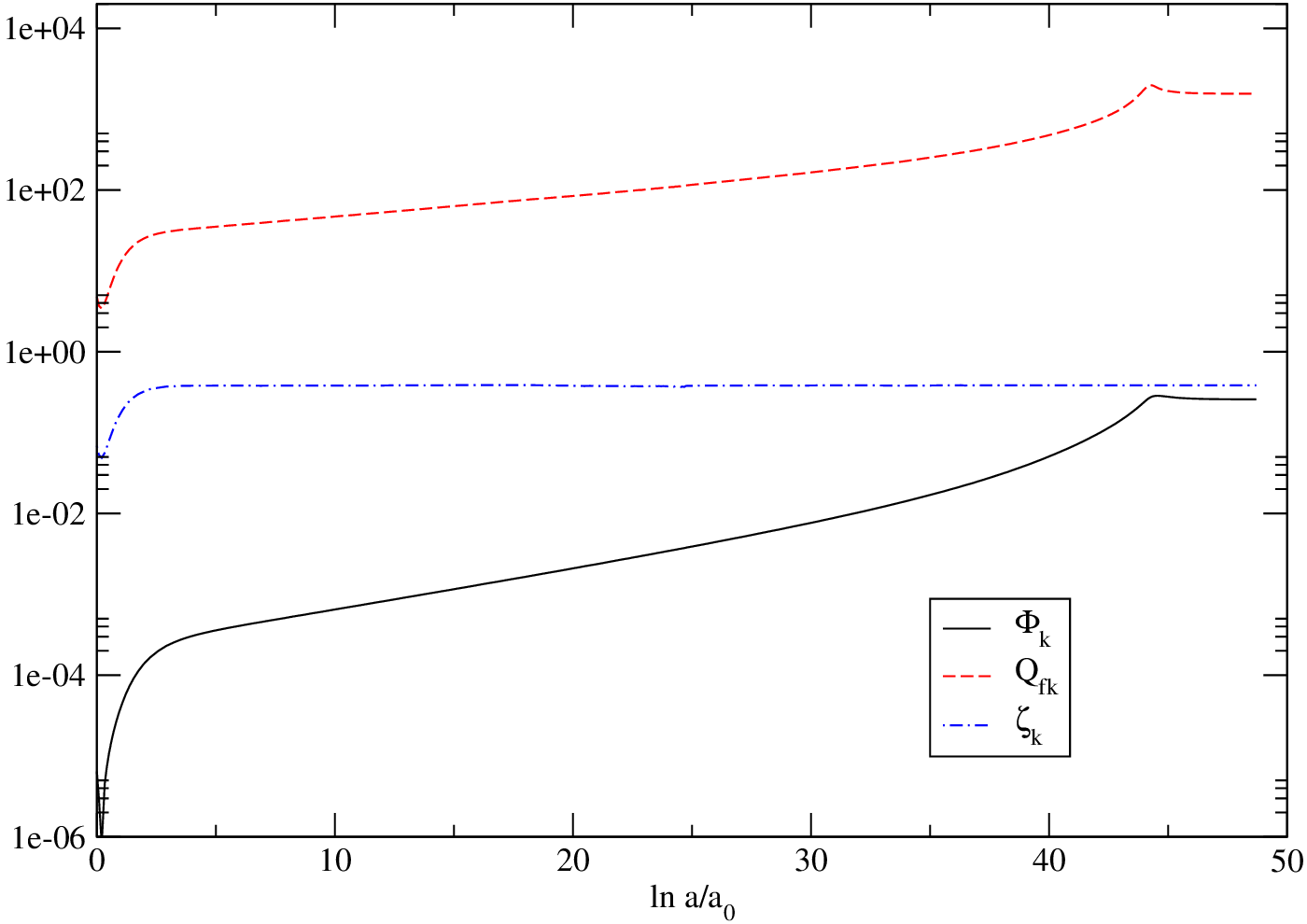}
\caption{Evolution of the Sasaki-Mukhanov variable $Q_{fk}$ in the
case when matter is a single NC fluid. Also shown are the 
curvature perturbation $\zeta_k$ and the Bardeen potential $\Phi_k$.
The simulations starts during the inflationary period and the results
shown are for the values $\lambda^{-1} = 10^{-4} m_{pl}$ and $\alpha = 1.099$.}
\label{fig_qf}
\end{figure}

We performed numerical calculations of the evolution of the linear
cosmological perturbations during  and after NC inflation. The 
simulations were done using the value $k = 5 a_i H_i$ for the comoving
momentum (the subscripts $i$ standing for the initial time of the analysis)
and started from $\lambda T_i = 9900$ to satisfy the condition
of Eq. (\ref{efold}).
Thus, the modes are inside the Hubble radius in the initial stage of 
the inflationary period. Quantum vacuum initial conditions for the
fluctuations were used \footnote{Since we are here interested in
the conditions under which the curvature fluctuations do not change
outside the Hubble radius, we can use vacuum initial conditions instead
of the thermal initial conditions which NC inflation \cite{Joao1,Joao2}
predicts.}. To implement these initial conditions, we postulate
the usual harmonic oscillator vacuum initial conditions for the
Fourier modes of the variable $v$ in terms of which the action for
fluctuations has canonical kinetic term. This variable is given by
\bea
v_k \, = \, \frac{a}{c_s}Q_{fk} \, = \, - z\zeta_k, \,
\eea
where the variable $z$ is
\bea
z \, = \, \frac{a\sqrt{\rho+{\cal P}}}{Hc_s} \, .
\eea
The variable $v_k$ obeys the following wave equation
\bea
v_k^{\prime\prime} + \left(k^2 c_s^2 -\frac{z^{\prime\prime}}{z}\right) v_k 
\, = \, 0,
\label{v_eq}
\eea
where a prime denotes the derivative with respect to conformal time $\eta$
($dt = a d\eta$). The initial conditions for the $k$'th Fourier mode
$v_k$ of $v$ are given by 
\bea
v_k(t_i) \, = \, \frac{1}{\sqrt{2k c_s}}, 
\quad \dot{v_k} (t_i) \, = \, -i\sqrt{\frac{k c_s}{2}}.
\eea
Then, from Eqs. (\ref{def_zeta}) and (\ref{eq_zeta}), we obtain the
initial conditions for $\Phi$ (suppressing the index $k$ for now)
\bea
& &\Phi_i \, = \, -\frac{\dot{H}}{H}\frac{a^2}{k^2 c_s^2}
\left(\frac{v_i}{z}\right)^{\cdot} \, = \, 4\pi G \frac{H z^2}{k^2}
\left(\frac{v_i}{z}\right)^{\cdot}, \\
& &\frac{H}{a}\left(\frac{a}{H}\Phi_i\right)^{\cdot}
\,  = \, \frac{\dot{H}}{H}\frac{v_i}{z} \, = \, 
-4\pi G \frac{\rho(1+w)}{H z} v,
\label{intial_bardeen}
\eea
where $v_i = v(t_i)$.

We set the cutoff mass $\lambda^{-1} = 10^{-4} m_{pl}$ 
to be consistent with the observational data \cite{Koh,Joao2}
and use $\alpha = 1.099$, a value which gives an inflationary
background. With the initial conditions described above
and these parameter values, we numerically solve the Eq. (\ref{eq_bardeen}).

As shown in Section 2, the speed of sound is given by $c_s^2 = c_a^2$.
Thus, during the inflationary phase $c_s^2$ is negative. This leads
to an instability of short wavelength fluctuations. Since the
focus of this paper is on the evolution of the fluctuations on
super-Hubble scales, and since - 
according to the prescription of \cite{Joao2,Koh} - we should
impose initial conditions for the fluctuations at the typical
thermal wavelength, and modes thus do not spend a long time on sub-Hubble
scales, we will not worry about this instability in this paper.
 
In Fig. \ref{fig_qf}, we display the evolution of $\Phi_k, Q_{fk}$ 
and $\zeta_k$ during and after inflation.
Fig. \ref{fig_qf} shows that the curvature perturbation $\zeta_k$ stays 
constant for super-Hubble scales 
both during inflation and after the end of inflation. This is
the same behavior as holds for adiabatic fluctuations in scalar field-driven
inflationary models.
Meanwhile, the time-dependent equation of state causes the Bardeen
potential $\Phi_k$ and also Sasaki-Mukhanov variable $Q_{fk}$
to increase for super-Hubble scales during inflation 
as time increases. We wish to emphasize that, 
as expected, there is no amplification of the
amplitude of curvature perturbations at the end of inflation.
This can be understood from the fact
that in Eq. (\ref{v_eq}), the function $z^{\prime\prime}/z$ term does 
not include any oscillating terms at the end of inflation. 

Our numerical analysis (Fig. \ref{fig_qf})
also shows that the curvature perturbation 
$\zeta_k$ as well as $\Phi_k$ and $Q_{fk}$ increase on
sub-Hubble scales . This is unlike what happens 
in the usual scalar field-driven inflation model. The difference,
as discussed above, is due to the fact that in our case the
sound speed $c_s^2$ is negative during inflation.

\section{Evolution of noncommutative fluid perturbations including an 
oscillating scalar field}

Particle physics models beyond the Standard Model, in particular those
based on supersymmetry, supergravity and superstring theory, typically
contain many scalar fields. These fields will be displaced from the
minima of their potentials by quantum fluctuations during the period
of inflation. After inflation, they will start to oscillate about
their minima. In this section we will study the effects of these
oscillations on the parametric excitation of entropy fluctuations on
super-Hubble scales.

In different contexts, cosmological perturbations in models with 
both scalar fields and fluids have been studied before. For example,
in warm inflation models \cite{Berera}, the scalar inflaton field is
coupled to a thermal fluid. Since the
scalar field evolution is typically overdamped, there will be
no parametric resonance effects. The growth of fluctuations in this context
has been considered e.g. in \cite{diMarco}.  Our situation is different
in that the dynamics is driven by the fluid rather than by the scalar
field. In a more general context, perturbations in a system consisting
of a scalar field plus an ideal gas were recently also considered in
\cite{Bartolo}, in particular with applications to quintessence cosmology
in mind.

In the following, we first discuss the equations of motion for the
cosmological fluctuations, and then present our numerical results.

\subsection{Equations of motion}

The background scalar field $\phi$ satisfies the Klein-Gordon equation
\bea
\ddot{\phi} + 3 H\dot{\phi} + V_{\phi} \, = \, 0,
\label{eq_bfield}
\eea
where $V_{\phi} = \partial V/\partial \phi$. In the presence
of both a fluid and a scalar field, the background Friedmann equation is
\bea
H^2 \, = \, \frac{8\pi G}{3} \left(\rho_f + \frac{1}{2}\dot{\phi}^2 
+V(\phi)\right).
\eea
where subscript $f$ denotes a fluid quantity.

\begin{figure}
\includegraphics[height=8cm]{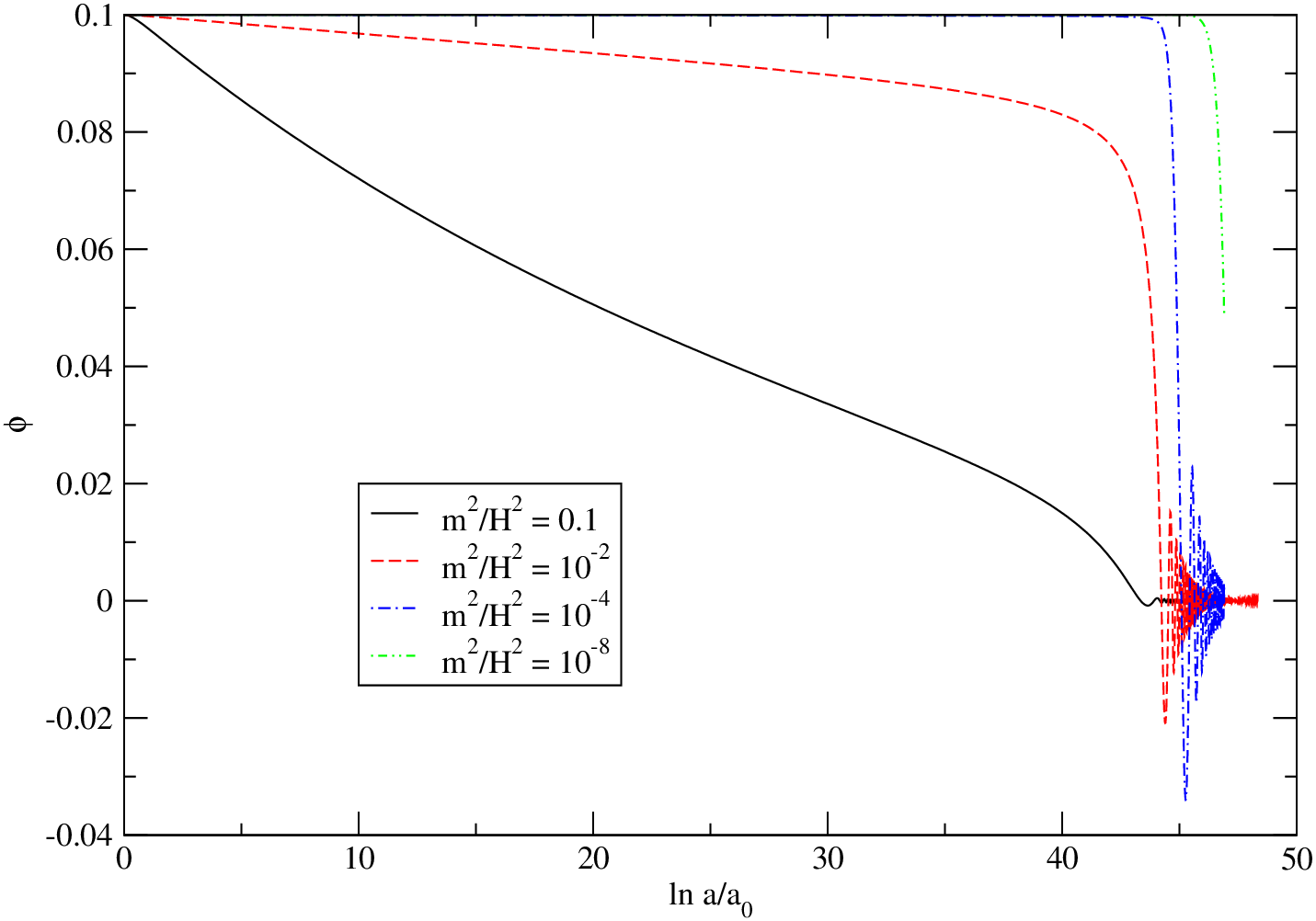}
\caption{Evolution of the scalar field during and after the period of 
NC inflation for various values of the scalar field mass $m$. The
parameters $\lambda^{-1} = 10^{-4} m_{pl}$ and $\alpha = 1.099$ were used.}
\label{fig_field}
\end{figure}

In Fig. \ref{fig_field}, we plot the evolution of a free scalar field
$\phi$ in a universe dominated by non-commutative radiation. Different
plots are for different values of the scalar field mass $m$.

The total perturbed energy density, pressure and momentum are
\bea
& &\delta \rho \, = \, \delta \rho_f + \delta \rho_{\phi}, \quad
\delta {\cal P} \, = \, \delta {\cal P}_f + \delta {\cal P}_f, \nonumber \\
& & \delta q \, = \, \delta q_f - \dot{\phi} \delta \phi,
\eea
where
\bea
\delta \rho_{\phi} \, &=& \, \dot{\phi}\delta \dot{\phi} -\dot{\phi}^2
\Phi +V_{\phi} \delta \phi, \\
\delta {\cal P}_{\phi} 
\, &=& \, \dot{\phi} \delta \dot{\phi} - \dot{\phi}^2 \Phi
-V_{\phi} \delta \phi.
\eea

By inserting the metric including cosmological fluctuations into
the curved space-time Klein-Gordon equations, one obtains the
following equation for the fluctuation of the scalar field
\bea
\delta \ddot{\phi}_k + 3 H\delta \dot{\phi}_k +\left(\frac{k^2}{a^2}
+V_{\phi\phi}\right)\delta \phi_k \, =
\, 4\dot{\Phi}_k \dot{\phi} - 2V_{\phi} \Phi_k \, .
\label{eq_field}
\eea
By subtracting $c_s^2$ times the $0-0$ perturbed Einstein equation
from the space-space perturbed Einstein equation, one obtains the
following evolution equation for the Bardeen potential
\bea
\ddot{\Phi}_k + (4+3c_s^2) H\dot{\Phi}_k 
&+& \left[\frac{k^2 c_s^2}{a^2} +2 \dot{H} + 3H^2 (1+c_s^2)
+4\pi G(1-c_s^2)\dot{\phi}^2 \right] \Phi_k       \nonumber \\
&=& \, 4\pi G[(1-c_s^2)\dot{\phi} \delta \dot{\phi}_k
-(1+c_s^2) V_{\phi} \delta \phi_k],
\label{eq_bardeen2}
\eea
where
\bea
\dot{H} = -4\pi G (\rho_f (1+w_f) +\dot{\phi}^2).
\eea
Note that in the above $c_s^2$ is the speed of sound of the fluid component.

The isocurvature perturbation for a multi-component system (the components
being labelled by indices $i, j, ...$) takes the form \cite{KS}
\bea \label{iso}
p\Gamma \, = \, \sum_i (c_{si}^2 - c_{ai}^2)\delta \rho_i
+\frac{1}{2}\sum_{i,j}\frac{h_i h_j}{h}(c_{ai}^2-c_{aj}^2)S_{ij},
\eea
where we have used the abbreviations
\bea
h \, = \, \rho+{\cal P}, \quad h_i \, = \, \rho_i + {\cal P}_i \, .
\eea
The first term in Eq. (\ref{iso}) represents the non-adiabatic pressures 
of the individual components and the second term the relative isocurvature 
perturbations between the different components.
If energy transfer between the components is neglected, $S_{ij}$ becomes
\bea
S_{ij} \, = \, \frac{\delta \rho_i}{\rho_i+{\cal P}_i} -
\frac{\delta \rho_j}{\rho_j+{\cal P}_j}.
\eea

In our case we consider one fluid and one scalar field, and the
relevant term, the expression for $S_{f\phi}$ is
\bea
S_{f\phi} \, &=& \, -\frac{3H^2\Phi_k+3H\dot{\Phi}_k+\frac{k^2}{a^2}\Phi_k}
{4\pi G\rho_f(1+w_f)} 
-\frac{\rho_f(1+w_f)+\dot{\phi}^2}{\rho_f (1+w_f) \dot{\phi}^2}
\delta \rho_{\phi}  \nonumber \\
&=& \, -\frac{1}{c_s^2\sqrt{\rho_f (1+w)}}
\left(\dot{Q}_f+\frac{3}{2}H(1-c_a^2)Q_f\right)
-\frac{1}{\dot{\phi}^2} (\dot{\phi}\dot{Q}_{\phi}
+V_{\phi}Q_{\phi}) \nonumber \\
& & \, -\frac{4\pi G}{H}\frac{1-c_s^2}{c_s^2}
(\dot{\phi}Q_{\phi} -\sqrt{\rho_f (1+w)}Q_f) \, .
\eea
Then, from Eq. (\ref{iso}), the relative isocurvature perturbation is
\bea \label{gamma1}
p\Gamma_{rel} \, = \, \frac{1}{2}\frac{\rho_f(1+w_f)\dot{\phi}^2}{
\rho_f(1+w_f)+\dot{\phi}^2}(c_{af}^2-c_{a\phi}^2)S_{f\phi},
\eea
where the speed of sound of the scalar field component is
\bea
c_{a\phi}^2 \, = \, \frac{\dot{p}_{\phi}}{\dot{\rho}_{\phi}}
=1+\frac{2V_{\phi}}{3H\dot{\phi}}.
\eea
As we have discussed in Section 2, there is no non-adiabatic pressure
for the NC fluid. The non-adiabatic pressure for the scalar field
can be written as
\bea \label{gamma2}
p\Gamma_{int} \, &=& \, -\frac{2V_{\phi}}{3H\dot{\phi}} \delta \rho_{\phi}\\
&=& \, -\frac{2 V_{\phi}}{3H\dot{\phi}}(\dot{\phi}\dot{Q}_{\phi}
+V_{\phi}Q_{\phi})+\frac{V_{\phi}\dot{\phi}}{\rho}
(\dot{\phi}Q_{\phi}-\sqrt{\rho_f(1+w_f)}Q_f). \nonumber
\eea
The sum of Eqs. (\ref{gamma1}) and (\ref{gamma2}) gives the total 
non-adiabatic pressure which is the source term in the equation 
of motion for the curvature perturbation $\zeta$ (see Eq. (\ref{eq_zeta})).

In the case of a multi-component system, the variable $\zeta$
which geometrically represents the curvature perturbation on the uniform 
energy density slices is given by \cite{Lyth03}
\bea
\zeta \, &=& \, 
\frac{\sum_i\dot{\rho}_i \zeta_i}{\sum_i \dot{\rho}_i} \\
&=& \, H\frac{-\sqrt{\rho_f(1+w_f)}Q_f+\dot{\phi}Q_{\phi}}{\rho_f(1+w_f)
+\dot{\phi}^2} \, , \nonumber
\eea
and its equation of motion is given by the generalization of
(\ref{eq_zeta}) as
\bea
\dot{\zeta}_k \, = \, \frac{H}{\dot{H}}\frac{k^2 c_s^2}{a^2}\Phi_k
-\frac{H}{\rho_f(1+w_f)+\dot{\phi}^2}p\Gamma
\label{eq_zeta2}
\eea
From this, we see that in the case of the NC fluid coupled to a scalar
field, there is a non-vanishing source for $\zeta_k$ which is not
suppressed on large scales (the second term on the right hand side
of Eq. (\ref{eq_zeta2})), and if the scalar field is oscillating, there
is a possibility of parametric resonance enhancement of $\zeta_k$.
We will now turn to a numerical study of this question.

\subsection{Numerical Calculations}
\begin{figure}
\includegraphics[height=8cm]{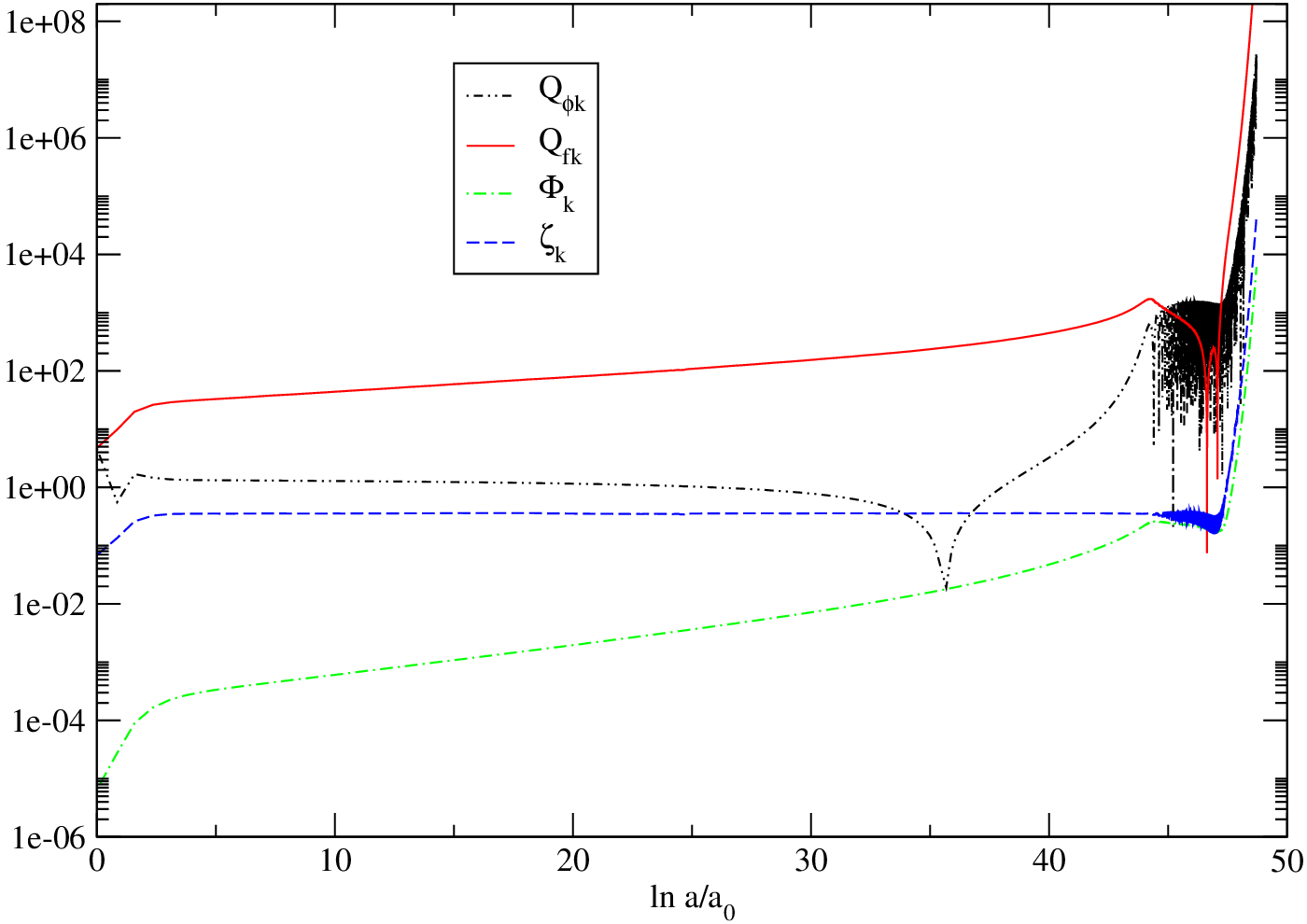}
\caption{Evolution of the individual $Q$ variables, namely $Q_{fk}$ for the 
NC fluid and $Q_{\phi k}$ for the scalar field, the Bardeen potential $\Phi_k$,
and curvature perturbation $\zeta_k$, in our model of NC inflation.
The results are for the choice of the NC inflation parameters 
$\lambda^{-1} = 10^{-4} m_{pl}$ and $\alpha = 1.099$. In the scalar field
sector, we set the initial value of $\phi$ as $\phi_i = 0.1 m_{pl}$,
$\dot{\phi}_i = 0$, and $m^2/H^2 = 10^{-2}$. 
}
\label{fig_qfs2}
\end{figure}
\begin{figure}
\includegraphics[height=8cm]{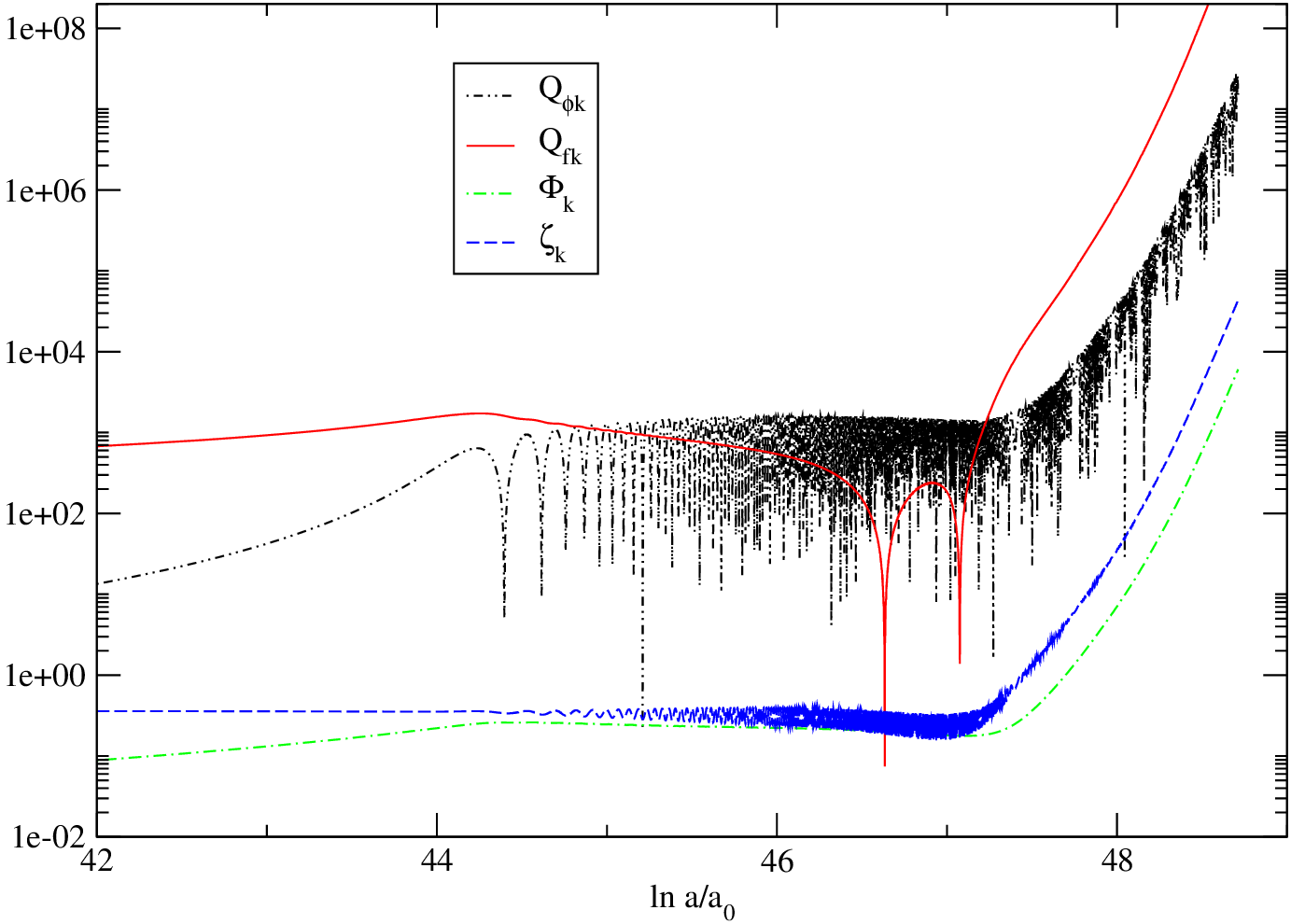}
\caption{A closer look at the region of oscillations in the simulation
of Fig. \ref{fig_qfs2}. 
}
\label{fig_qfs2-1}
\end{figure}
\begin{figure}
\includegraphics[height=8cm]{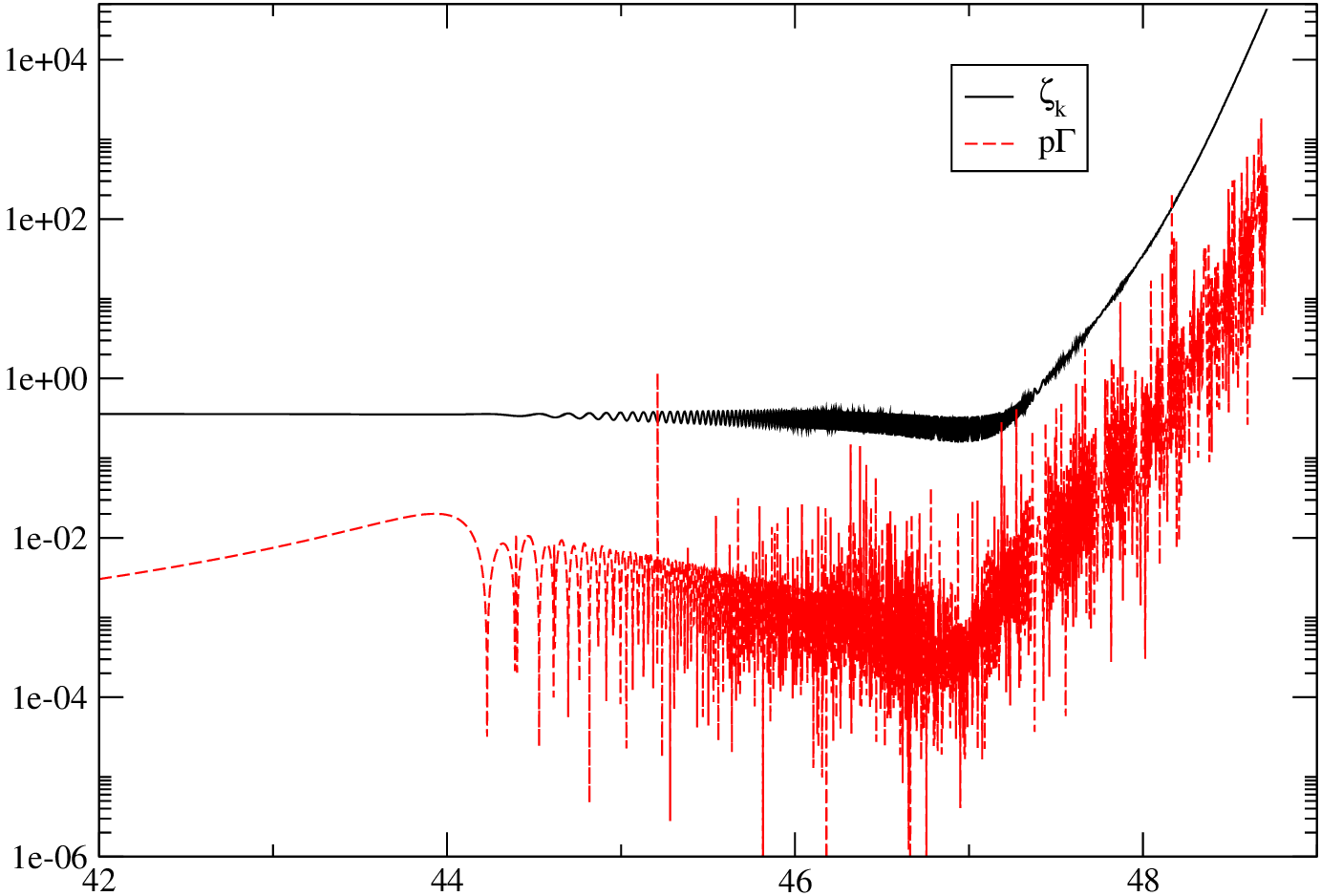}
\caption{Comparison of the curvature perturbation $\zeta_k$ and the
isocurvature fluctuation $p\Gamma$ in the simulation of Fig. \ref{fig_qfs2}.
}
\label{fig_iso2}
\end{figure}

We have evolved Eqs. (\ref{eq_field}) and (\ref{eq_bardeen2})
numerically for a massive scalar field potential
\bea
V(\phi) \, = \, \frac{1}{2} m^2 \phi^2.
\eea
The simulations began when the perturbation modes are still inside 
the horizon, and the scalar field fluctuations were started in
their vacuum state given by
\bea
\delta \phi_k(t_i) \, &=& \, \frac{1}{\sqrt{2k}}, 
\quad 
\delta \dot{\phi}_k(t_i) \,  = \, -i\sqrt{\frac{k}{2}}.
\eea
Since we are considering initial conditions for which the energy
density is dominated by the NC fluid, i.e. $\rho_f \gg \rho_{\phi}$ during the
early stages of the inflation,  we used the same initial conditions
for $\Phi$ as in Eq. (\ref{intial_bardeen}).

In Fig. \ref{fig_field} we plot the evolution of the background scalar field 
during and after inflation for different values of the mass parameter of 
the scalar field. We used the NC fluid parameters
$\lambda^{-1} = 10^{-4} m_{pl}$ and $\alpha=1.099$, and 
chose the initial value of the background scalar field as 
$\phi_i = 10^{-1}m_{pl}$  with vanishing initial velocity. We considered
different values of the mass. 

For $m > H$, the scalar field oscillations
are rapidly damped during inflation, and 
no oscillations arise during the post-inflationary era. For $m \ll H$,
the scalar field is over-damped during and immediately after inflation
and thus hardly rolls at all during the time interval of our simulations. 
The interesting case for our purpose arises when the mass $m$ is 
slightly but not much smaller than $H$. In this case, the  
scalar field rolls slowly during inflation 
and, when the inflationary phase ends and the Hubble parameter starts
to decrease, the slowly decreasing Hubble parameter 
becomes comparable in value to the mass of the scalar field and then 
begins to oscillate about the minimum of its potential. The time scale
of the oscillations becomes smaller than the Hubble time as $H$
decreases. This provides the conditions where parametric resonance
excitation of the entropy modes is expected \cite{Fabio2}.

The evolution of $Q_f, Q_{\phi}, \Phi_k$ and $\zeta_k$ are plotted
in Fig. \ref{fig_qfs2} for $m^2 = 10^{-2} H^2$ and $\phi_i = 10^{-1} m_{pl}$
and for the same parameter values
of $\lambda$ and$\alpha$  as in Fig. \ref{fig_field}.
With these parameter values, parametric amplification of the curvature
perturbation results after the end of the inflationary phase.

To see the region in which parametric resonance occurs in detail, 
in Fig. \ref{fig_qfs2-1} we focus on the time interval where the scalar
field oscillations occur. For a value of the mass given
by $m^2 = 10^{-2} H^2$, the 
oscillation of the background scalar field during this time
interval are seen (see Fig. \ref{fig_field}) to cause oscillations
of the scalar field fluctuation variable $Q_{\phi}$. Via the
gravitational coupling of the two matter components, this leads to 
the parametric amplification of the fluid fluctuation
variable $Q_f$. 
 
To understand the parametric resonance after inflation in our NC
fluid model of inflation with an oscillating scalar matter field,
 we must consider the coupled system of evolution
equations for $Q_f$ and $Q_{\phi}$ which are given in the Appendix
(Eqs. (\ref{qeq_s}) and (\ref{qeq_ncs})). We assume that the time scale 
of the resonance is short compared to the Hubble time $H^{-1}$. 
In this case, we can neglect that Hubble damping terms in the evolution
equations. We will also drop the other terms involving only first
derivatives of the $Q$ variables - we later check the self-consistency
of this approximation - and we will also drop coefficients
of $Q_f$ and $Q_{\phi}$ terms which are obviously
suppressed compared to other coefficients. 
Then, we obtain approximate equations 
\bea
\ddot{Q}_{fk} + \left[\frac{k^2 c_s^2}{a^2}
+ 6 \pi G A \rho_f - 6 \pi G B \dot{\phi}^2 \right] Q_{fk} 
- 4 \pi G {\sqrt{\rho_f (1 + w_f)}} 
[C + 3 \dot{\phi} D] Q_{\phi k} 
\, \simeq \, 0 \, ,
\label{qeq_ncs2}
\eea
where 
\bea
A \, &\equiv& \, (1+w)(1+3c_s^2)+(1-c_s^2)(1+c_s^2)-2(1+w_f)^2
\frac{\rho_f}{\rho}, \\
B \, &\equiv& \, -c_s^2(1-c_s^2)+(1+w)(1+c_s^2)\frac{\rho_f}{\rho}, \\
C \, &\equiv& \, (1 + c_s^2) \frac{V_{\phi}}{H},\\
D \, &\equiv& (1 + c_s^2) - (1 + w) \frac{\rho_f}{\rho}\, ,
\eea
and 
\bea
\ddot{Q}_{\phi k} + \left[\frac{k^2}{a^2} + V_{\phi \phi}
+ 16 \pi G V_{\phi} \frac{\dot \phi}{H} \right] Q_{\phi k} 
+ 8 \pi G \sqrt{\rho_f(1+w_f)} 
\left[\frac{V_{\phi}}{H} + \epsilon {\dot \phi} \right] Q_{f k} 
\, \simeq \, 0 \, ,
\label{qeq_ncs3}
\eea
where
\bea
\epsilon = 3+\frac{3}{4}\frac{(1-c_s^2)^2}{c_s^2}-\frac{3}{4}
\frac{(1+c_s^2)(1+w_f)}{c_s^2}\frac{\rho_f}{\rho}.
\eea
Note that at the end of the inflationary phase, 
$\rho_f/\rho \sim 1/2$ because of $\rho_f \simeq \rho_{\phi}$.
This system of equations represents two coupled harmonic oscillators with
mass terms which, during the time interval when $\phi$ is oscillating,
contain some periodically varying coefficients. We
can diagonalize this system of equations to get two decoupled equations
of Mathieu type \cite{TB,KLS2}. 
\bea
\frac{d^2 Q_k}{dz^2} +[A(k) - 2q \cos 2z] Q_k \, = \, 0 \, ,
\eea
where $Q$ stands for one of the eigenvectors of the diagonalization process
and $z = mt$.

We need to estimate the values of $A(k)$ and $q$ once $\phi$ begins to 
oscillate and can be approximated as $\phi = \phi_e \sin mt$ where $\phi_e$ 
is the amplitude of the scalar field at the end of inflation.
While for $q\ll 1$, the resonance occurs only in narrow bands \cite{TB},
for $q >1$, the resonance occurs for a broad range of values of $k$ 
\cite{KLS1}. Our numerical simulations indicate that we are in the
broad resonance region. Let us for a moment neglect the coupling terms
between the $Q_{f}$ and $Q_{\phi}$ equations. In this case, the
value of $q$ in the $Q_f$ equation would be of the order
\bea
q \, \sim \, 6 \pi \left(\frac{\phi_e}{m_{pl}}\right)^2
\eea
and would be less than $1$. However, if we estimate the amplitude $\tilde{q}$
of the periodically varying coefficient in the cross coupling term
in Eq. (\ref{qeq_ncs2}), we find
\bea \label{est2}
{\tilde{q}} \, \sim \, \frac{H}{m} \frac{\phi_e}{m_{pl}} \, . 
\eea
Inserting the value of $H$ at the beginning of inflation,
we obtain a number which is of order unity. Looking at the $Q_{\phi}$
equation we find a very similar result: the 
periodically varying coefficient multiplying $Q_f$ leads to a
value of $\tilde{q}$ given by Eq. (\ref{est2}) and hence appears
to contribute to the sourcing of the broad resonance. The problem
with this argument, however, is that the value of $H$ at the end
of inflation has decreased to a value smaller than $m$, and hence
the above estimate of $\tilde{q}$ in Eq. (\ref{est2}) 
gives a value smaller than 1.
Returning, however, to the general equations in the Appendix, we
see that there are coefficients which contribute to $q$ which are
enhanced by $c_s^{-1}$ and hence, during the period when the
equation of state transits from that of inflation to that of regular
radiation, leads to an enhancement of $q$. 

To summarize the discussion in the above paragraph: We have
presented analytical estimates which support the numerical results
which imply that the parametric amplification of the coupled $Q_f$ and
$Q_{\phi}$ system is in the broad resonance domain, as shown in
Fig. \ref{fig_qfs2-1}. Note that the $Q_f$ and $Q_{\phi}$ equations
have a singularity when $c_s^2 = 0$. However, we are numerically solving
not the $Q$ equations, but the $\Phi$ equation, and our evolution
equations are non-singular. 

The amplification of $Q_{\phi}$ and $Q_f$ also gives rise to
a parametric amplification of $\zeta$, as can be seen in Fig. \ref{fig_qfs2-1}.
From Eq. (\ref{eq_zeta2}), it can be seen that 
the large growth  of $\zeta$ is generated through the isocurvature
source term. We have compared the magnitudes of the 
curvature perturbation and the isocurvature 
source term in Fig. \ref{fig_iso2}. This confirms numerically that the 
amplification of $\zeta$ occurs once the
isocurvature source has become sufficiently large.

As a further consistency check, we can perform a rough estimate of the 
duration of the period of oscillations
until the amplitude of the curvature perturbation can start to grow.
As soon as the scalar field starts to oscillate, the isocurvature perturbation 
in Eq. (\ref{eq_zeta2}) grows exponentially as 
$\exp(\mu m(t-t_e))$ where $t_e$ is the time when the oscillation begins
and the Floquet exponent $\mu$ describes the rate of exponential growth 
of the instability.
For $q \gg 1$, unstable modes grows extremely rapidly with $\mu \sim 0.2$
\cite{KLS1}.
Then after integrating Eq. (\ref{eq_zeta2}), the curvature 
perturbation becomes
\bea
\zeta(t) \, \simeq \, \zeta(t_e) -\frac {H(t_e) p\Gamma(t_e)}
{\mu m(\rho_f(t_e)(1+w_f(t_e))
+\dot{\phi}(t_e))}e^{\mu m(t-t_e)} \, .
\eea

The time $t$ when the increase in the curvature perturbation
becomes visible compared to its initial value is given by
$\Delta \zeta \equiv \zeta(t) - \zeta(t_e)
\sim {\cal O}(1) \zeta(t_e)$ at the time when the amplitude
of the curvature perturbation begins to grow. Then we can estimate
the duration of the oscillatory period 
until the curvature perturbation starts to
increase substantially compared to their initial value
\bea
t-t_e \, \simeq \, \frac{1}{\mu m}\ln \left(3\mu m
\frac{\zeta(t_e)}{p\Gamma(t_e)}
\frac{m^2 \phi_e^2}{H(t_e)}\right)
\label{oscil_time}
\eea
where we have used $\rho_f(t_e) \simeq \rho_{\phi}(t_e),
\dot{\phi}(t_e) \sim m \phi(t_e)$, and $w_f(t_e) \sim {\cal O}(1)$.
$m \sim 10 H(t_e)$,  $\phi(t_e) \sim 10^{-1} m_{pl}$ 
and $\zeta(t_e)/p\Gamma(t_e) \sim 10^2$
at the beginning of oscillation
for the parameter values in Fig. \ref{fig_iso2}.
With these parameters, Eq. (\ref{oscil_time})
gives 
\bea
t - t_e \, \sim \, \frac{6}{\mu m} \, \sim \, 
6 \frac{1}{\mu} \frac{H(t_e)}{m} H^{-1}(t_e) \, \sim \, 3 H^{-1}(t_e) \, .
\eea

From the numerical calculations (Fig. \ref{fig_iso2}),  
$(t-t_e)_{numeric} \simeq \Delta \ln (a/a_0) H^{-1}(t_e) \sim 3 H^{-1}(t_e)$.
This is consistent with the above analytical estimate.

\begin{figure}
\includegraphics[height=8cm]{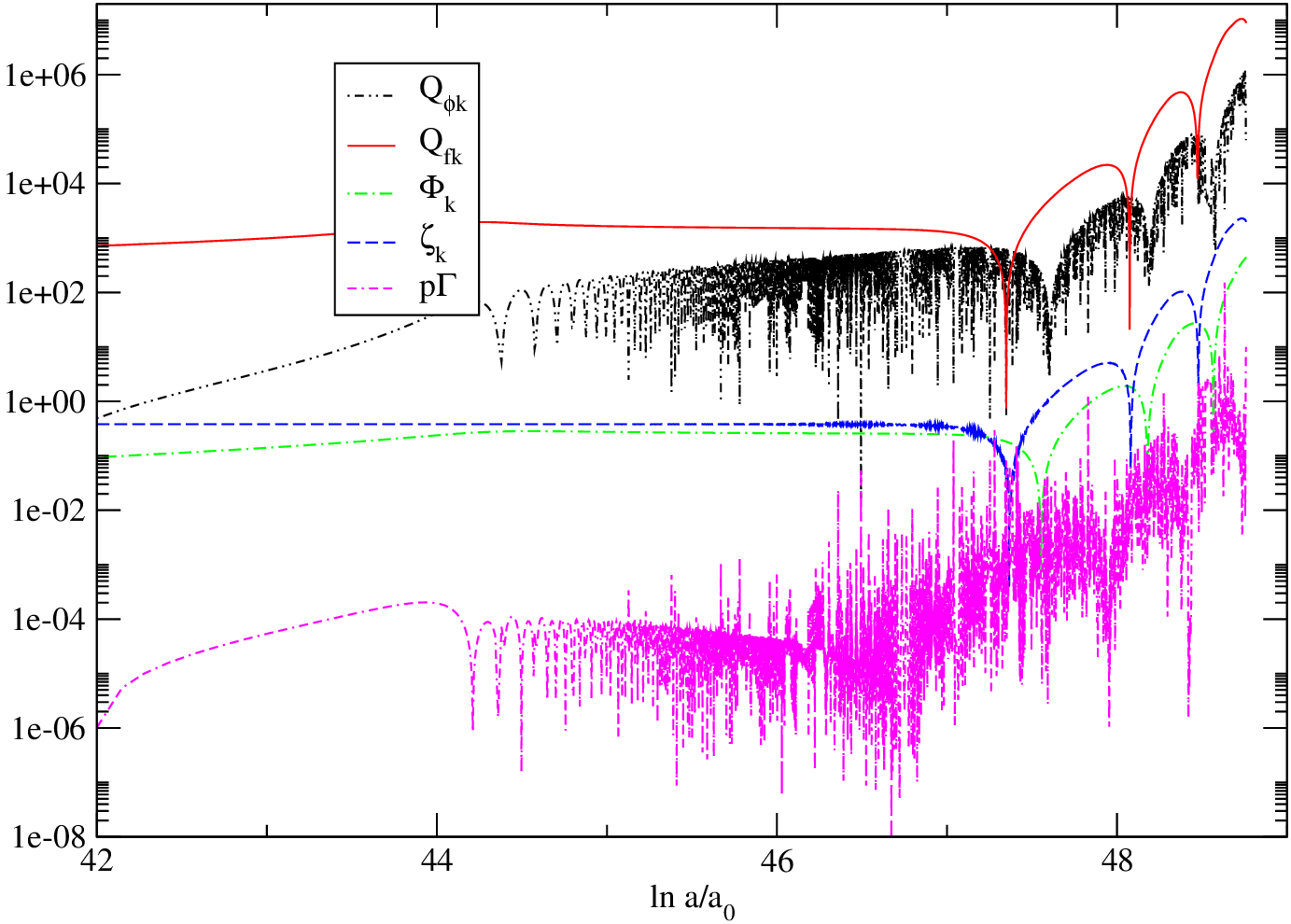}
\caption{Evolution of $Q_{fk}$ for the NC fluid, $Q_{\phi k}$
for the scalar field, the Bardeen potential $\Phi_k$,
the curvature perturbation $\zeta_k$, and $p\Gamma$
during the inflationary period for the parameter values
$\lambda^{-1} = 10^{-4} m_{pl}$ and $\alpha = 1.099$. 
We set $\phi_i = 10^{-2} m_{pl}$,
$\dot{\phi}_i = 0$, and $m^2/H^2 = 10^{-2}$. 
}
\label{fig_qfs3}
\end{figure}

We end our discussion by showing the results of a simulation
with different values of the parameters. In Fig. \ref{fig_qfs3},
we plot $Q_{fk},Q_{\phi k}, \Phi_k, \zeta_k$ and $p\Gamma$ 
for $\phi_i = 10^{-2}m_{pl}$ and $m^2/H^2 =10^{-2}$. Once again
we find parametric amplification of the curvature perturbation.

\section{Discussion}

In this paper we have studied the evolution of cosmological
fluctuations through the transition between inflationary phase
and radiation phase in our non-commutative inflation model
\cite{Joao2,Koh}. In this model, there is no traditional
reheating phase with oscillating inflaton field. Thus, if
there is no matter entropy mode which oscillates, there is
no possibility for a resonance growth of curvature fluctuations
on super-Hubble scales. As we show here, if a scalar matter field
happens to be oscillating at the end of inflation, which is
expected to happen in many models beyond the Standard Model of
particle physics with light moduli fields (with mass of the
order of the Hubble constant during the inflationary phase),
then there can be parametric resonance of curvature fluctuations
at the end of inflation on super-Hubble scales. This resonant growth 
happens provided that the scalar field
mass at the end of the period of inflation is comparable to the
Hubble rate at that time.

The resonance we study here is different from the usual
resonance in scalar field-driven models, in which the resonance
is driven by the coherent oscillations of the dominant component
of matter. In our situation, it is oscillations of a sub-dominant
mode which is sourcing the resonance of the curvature fluctuations.
We show that there is an entropy mode which undergoes resonance,
and once the entropy mode has acquired a sufficiently large
amplitude, a resonant growth of the curvature fluctuations is
induced. 

A characteristic feature of our non-commutative inflation model
is the fact that the fluid speed of sound $c_s^2$ is negative
during the phase of accelerated expansion. This appears to lead
to an instability of fluctuations on small scales. Since this work
focuses on the transition of fluctuations through the period when the
equation of state of the background changes, rather than on the
question of initial conditions for the inflationary phase, we
here to not address this problem. However, we will need to come back
to this issue in future work. 

\acknowledgments

SK was supported by the Korea Research Foundation Grand funded
by the Korean Government (MOEHRD; KRF-2006-214-C00013). The
research of RB is supported by an NSERC Discovery Grant, by
funds from the Canada Research Chairs Program, and by funds from
a FQRNT Team Grant. We wish to thank J. Magueijo for interesting
discussions.

\appendix
\section{Equations for $Q_f$ and $Q_{\phi}$ for a model
consisting of one matter fluid and one scalar field}

The Sasaki-Mukhanov variables for the fluid and the scalar field are
defined in Eqs. (\ref{def_qf}) and (\ref{def_qphi}), respectively.
They satisfy the following set of coupled equations:
\bea
& & \ddot{Q}_{\phi k} + 3H\dot{Q}_{\phi k}
-4\pi G\frac{\dot{\phi}}{H c_{sf}^2}(1-c_{sf}^2) \sqrt{\rho_f(1 +w_f)} 
\dot{Q}_{fk} 
 +\biggl[ \frac{k^2}{a^2} + V_{\phi\phi}
+16\pi G\frac{\dot{\phi}}{H}V_{\phi} \nonumber \\
& &~~~~- 16\pi^2 G^2
\frac{\dot{\phi}^2(1-c_{sf}^2)}{H^2 c_{sf}^2}\rho_f(1+w_f)
-32\pi^2 G^2\frac{\dot{\phi}^2}{H^2}\rho_f(1+w_f)
-32\pi^2 G^2 \frac{\dot{\phi}^4}{H^2} \nonumber \\
& &~~~~+24\pi G\dot{\phi}^2\biggr] Q_{\phi k}
-\biggl[ 8\pi G\frac{V_{\phi}}{H}+6\pi G\dot{\phi}
\frac{(1-c_{sf}^2)^2}{c_{sf}^2}+24\pi G\dot{\phi} \nonumber \\
& &~~~~-16\pi^2 G^2 \frac{\dot{\phi} (1+c_{sf}^2)}{H^2 c_{sf}^2}\rho_f(1+w_f)
-32\pi^2 G^2\frac{\dot{\phi}^3}{H^2}\biggr] \sqrt{\rho_f(1+w_f)}
Q_{fk} \, = \, 0, 
\label{qeq_s}\\
& & \ddot{Q}_{fk} +\left[ 3H -\frac{\dot{c}_{sf}^2}{c_{sf}^2}\right] 
\dot{Q}_{fk} 
+4\pi G\frac{\dot{\phi}}{H}(1-c_{sf}^2)\sqrt{\rho_f(1+w_f)}
\dot{Q}_{\phi k} + \biggl[\frac{k^2 c_{sf}^2}{a^2}
+6\pi G\rho_f(1+w_f)(1+3c_{sf}^2) \nonumber \\
& &~~~~ - 6\pi G\dot{\phi}^2(1-c_{sf}^2)
+\frac{9}{4}H^2 (1-c_{sf}^2)(1+c_{sf}^2)
-32\pi^2 G^2 \frac{\rho_f^2(1+w_f)^2}{H^2} \nonumber \\
& &~~~~-16\pi^2 G^2 \frac{\dot{\phi}^2}{H^2}\rho_f(1+w_f)(1+c_{sf}^2)
-\frac{3}{2}H\dot{c}_{sf}^2-\frac{3}{2}H(1-c_{sf}^2)
\frac{\dot{c}_{sf}^2}{c_{sf}^2}+4\pi G\frac{\rho(1+w_f)}{H}
\frac{\dot{c}_{sf}^2}{c_{sf}^2}
\biggr] Q_{fk}   \nonumber \\
& &~~~~-4\pi G\sqrt{\rho_f(1+w_f)}\biggl[ (1+c_{sf}^2)\frac{V_{\phi}}{H}
-8\pi G\frac{\dot{\phi}}{H^2}\rho_f(1+w_f)
+3\dot{\phi}(1+c_{sf}^2)
- 4\pi G\frac{\dot{\phi}^3}{H^2}(1+c_{sf}^2) \nonumber \\
& &~~~~+\frac{\dot{\phi}}{H}\frac{\dot{c}_{sf}^2}{c_{sf}^2}
\biggr] Q_{\phi k} \, = \, 0,
\label{qeq_ncs}
\eea
where we have used $c_{sf}^2 = c_{af}^2$.
Similar equations were derived in \cite{diMarco} for constant $w_f
=c_{af}^2 = c_{sf}^2$ and with a friction term.
Note that Eq.(\ref{qeq_ncs}) becomes a single fluid equation for $\phi =0$.
And Eq.(\ref{qeq_s}) agrees with the equation for a single scalar
field equation when $\rho_f = 0$.

\end{document}